\documentclass[12pt,a4paper]{article}
\usepackage{hyperref}
\usepackage{url}
\usepackage{amsthm}
\usepackage{amsmath, amssymb}

\newcommand{\id}[1]{\ensuremath{\mathrm{id}}}

\newcommand{\half}{\mbox{\footnotesize $\frac{1}{2}$}}

\newcommand{\er}{\eqref}
 
\newcommand{\beq}{\begin{equation}}
\newcommand{\eeq}{\end{equation}} 
\newcommand{\bea}{\begin{eqnarray}}
\newcommand{\eea}{\end{eqnarray}}

 \newcommand{\til}{\tilde}
\newcommand{\raw}{\rightarrow}

 \newcommand{\Raw}{\Rightarrow}

\newcommand{\x}{\times}

\newcommand{\cin}{C^{\infty}}

\newcommand{\al}{\alpha} 
 
\newcommand{\dl}{\delta}

\newcommand{\rh}{\rho} \newcommand{\sg}{\sigma}
\newcommand{\Sg}{\Sigma}  
 \newcommand{\phv}{\varphi}
\newcommand{\ch}{\ch} \newcommand{\ps}{\psi}

\newcommand{\inv}{^{-1}}

\newcommand{\Tr}{\mbox{\rm Tr}\,}

 \newcommand{\g}{\mathfrak{g}}

 \newcommand{\R}{{\mathbb R}}
 
  \makeatletter
 
\makeatletter
\def\moverlay{\mathpalette\mov@rlay}
\def\mov@rlay#1#2{\leavevmode\vtop{%
   \baselineskip\z@skip \lineskiplimit-\maxdimen
   \ialign{\hfil$\m@th#1##$\hfil\cr#2\crcr}}}
\newcommand{\charfusion}[3][\mathord]{
    #1{\ifx#1\mathop\vphantom{#2}\fi
        \mathpalette\mov@rlay{#2\cr#3}
      }
    \ifx#1\mathop\expandafter\displaylimits\fi}
\makeatother

\newcommand{\Mi}{\mathbb{M}}
\newcommand{\mghd}{\textsc{mghd}}

\newcommand{\pde}{\textsc{pde}}

\renewcommand{\g}{\tilde{g}}
\newcommand{\tn}{\tilde{\nabla}}
\newcommand{\ali}{\begin{align}}
\newcommand{\elin}{\end{align}}
\newcommand{\X}{\mathfrak{X}}

\newcommand{\n}{\nabla}

\renewcommand{\g}{\tilde{g}}

\newcommand{\GR}{{\sc gr}}

\newtheorem{definition}{Definition}

\newtheorem{theorem}[definition]{Theorem}
\newtheorem{proposition}[definition]{Proposition}

\newtheorem{corollary}[definition]{Corollary}
\newcommand{\HA}{Hole Argument}
\topmargin = - 1 cm \textheight = 23 cm \textwidth = 15 cm
\usepackage[symbol]{footmisc}
\renewcommand{\thefootnote}{\fnsymbol{footnote}}
\begin{document}
\pagenumbering{arabic} \setlength{\unitlength}{1cm}
\begin{center}
\begin{Large}
{\bf Reopening the Hole Argument}\footnote[1]{This paper is a tribute to the weekly Cambridge--LSE \emph{Philosophy of Physics Bootcamp}, in which the \HA\  has often been discussed. I am  indebted to the organizers, Jeremy Butterfield and Bryan Roberts, as well as to
Henrique Gomes, Hans Halvorson, Joanna Luc,  and JB Manchak for comments on earlier drafts. Three anonymous referees also made major contributions to this final version by their suggestions and their criticism. Finally, I  wish to thank Michel Janssen for historical comments related to the Introduction of this paper.
}
\end{Large}
\medskip

\begin{large}
 Klaas Landsman\vspace{5mm}
 \end{large}

 Department of Mathematics and  Radboud Center for Natural Philosophy\\
   Radboud University, Nijmegen, The Netherlands.
Email:
\texttt{landsman@math.ru.nl}

 \begin{abstract} 
\noindent This expository paper relates the Hole Argument in  general relativity (\GR) to the well-known theorem of Choquet-Bruhat and  Geroch  (1969) on the existence and uniqueness of globally hyperbolic solutions to the Einstein field equations. Like the Earman--Norton (1987) version of the 
 Hole Argument (which is originally due to Einstein), this theorem exposes the tension beween determinism and some version of spacetime substantivalism. But it seems less vulnerable to the campaign by Weatherall (2018) and followers to close the Hole Argument on the basis of ``mathematical practice'', since the theorem only talks about isometries and hence does not make the pointwise identifications via diffeomorphisms that Weatherall objects to. Among other implications of the theorem  for the philosophy of \GR, we reconsider
  Butterfield's (1987) influential definition of determinism. This  should be amended if its goal is to express the idea that \GR\ is deterministic in the absence of Cauchy horizons, although its original form does capture the way \GR\  is indeterministic in their presence!  Furthermore, in \GR\ isometries come out as gauge symmetries, as do Poincar\'{e} transformations in special relativity. 
  
    Finally, I discuss some implications of the theorem for the philosophy of science: accepting the determinism horn still requires a choice between Frege-style abstractionism and Hilbert-style structuralism; and, within the latter, between structural realism and empiricist structuralism (which I favour). 
 \end{abstract}\end{center}
 
 \tableofcontents

\thispagestyle{empty}
\renewcommand{\thefootnote}{\arabic{footnote}}
\newpage
\section{Introduction}\label{intro} 
Initially, the \HA\ (\emph{Lochbetrachtung}) was an episode in Einstein's struggle between 1913--1915 to find the gravitational field equations of general relativity. At a time when he was already unable to find generally covariant  equations for the gravitational field (i.e.\ the metric) that had the correct Newtonian limit and satisfied energy-momentum conservation, the \HA\ confirmed him in at least temporarily giving up the idea of general covariance (which he later recovered without ever mentioning the \HA\ again at least in print).\footnote{See Janssen and  Renn (2022) for the final reconstruction of Einstein's struggle, with \S 4.1 devoted to the \HA.
The earliest known reference to the \HA\ is in a memo by Einstein's friend and colleague Besso  dated August 1913, provided this dating is correct (Janssen, 2007). Einstein subsequently presented his argument four times in print; I just cite Einstein (1914) as the paper containing his final version. Implicitly, his later point-coincidence argument (Einstein, 1916) was his own reply to his \HA\ (Norton, 1993; Giovanelli, 2021). 
See Stachel (2014),  Norton (2019),  Pooley (2022), and Gomes and  Butterfield (2023a), and references therein for reviews of the \HA\ in both a historical and a modern context. 
 } 
 Einstein's invention of the argument formed part of his analysis of the interplay between general relativity (of motion), general covariance (of  physical equations under coordinate transformations), and determinism (here: of the field equations of general relativity). 
 
Thus Einstein felt he had to choose between determinism and general covariance; the  recent emphasis on the  tension between determinism (siding with relationalism) and substantivalism is due to Earman and  Norton (1987). But since for Einstein the opposition between substantivalism and relationalism was closely related to the problem of absolute versus relative motion and hence to his putative ``principle of general relativity'' (Earman, 1989), he would certainly have been interested in it. 

 In modernized form (using a global perspective  and replacing Einstein's coordinate transformations  by diffeomorphisms), his reasoning was essentially as follows:\footnote{We write the Einstein tensor as $\mathrm{Ein}(g)$, where its dependence on the metric $g$ is explicitly denoted; in coordinates we have $\mathrm{Ein}(g)_{\mu\nu}=G_{\mu\nu}=R_{\mu\nu}-\half g_{\mu\nu}R$.}
\begin{itemize}
\item Let $(M,g)$ be a spacetime.\footnote{A spacetime is a smooth four-dimensional connected Lorentzian manifold with time orientation (this nomenclature of course hides philosophical issues to be discussed later in this paper).
 More generally, my notations and conventions follow Landsman (2021) and are standard, e.g.\ 
  spacetime indices are Greek whereas spatial ones are Latin, the metric has signature $-+++$, etc.
\label{4dst}}
The  transformation behaviour of the Einstein tensor $\mathrm{Ein}(g)$ under diffeomorphisms $\psi$ of the underlying manifold $M$ is
\begin{equation}
\psi^*(\mathrm{Ein}(g))=\mathrm{Ein}(\psi^*g).
\end{equation}
Similarly,  for any healthy energy-momentum tensor $T(g,F)$ constructed from the metric $g$ and the matter fields $F$ that matter we should have
\begin{equation}
\psi^*(T(g,F))=T(\psi^*g,\psi^*F).
\end{equation}
Consequently, if $g$ satisfies the Einstein equations $\mathrm{Ein}(g)=8\pi\, T(g,F)$, then $\psi^*g$ satisfies these equations for the transformed matter fields $\psi^*F$. 
\item Now consider an open connected vacuum region $H$ in spacetime, possibly surrounded by matter (i.e.\ $F=0$ in $H$); $H$ is referred to as a ``hole'', whence the name of the argument.\footnote{ Einstein's arrangement looks unnatural compared to Hilbert's (1917) reformulation as an initial-value problem in the \pde\ sense, see Proposition \ref{HHA} below, but Einstein was  inspired by Mach's principle, where ``fixed stars at infinity'' determine the local inertia of matter; see Maudlin (1990), Hofer (1994), and Stachel (2014). 
 An argument that actually favours Einstein's curious setting for the \HA\ is this: the smaller the hole, i.e.\ the larger the complement of the hole, the greater the challenge to determinism, for if even things almost everywhere except in a tiny hole fail to determine things inside that hole, then we should really worry (Butterfield, 1989).
This pull admittedly gets lost in the initial-value formulation of the argument  below.
See Muller (1995) for the explicit construction of a hole diffeomorphism (the only one I am aware of). 
 \label{EMach1}}
Furthermore, find a diffeomorphism $\psi$ that is nontrivial inside $H$ and equals the identity outside $H$, so that
in particular, 
\beq
T(\psi^*g,\psi^*F)=T(\psi^*g,F)=T(g,F),
\eeq
both outside $H$ (where $\psi$ is the identity) and inside $H$ (where $T(g,F)=0$). 
\item It follows from the previous  points that  if $g$ satisfies the Einstein equations for some  energy-momentum tensor $T$, then so does $\psi^*g$. Hence the spacetimes $(M,g)$ and $(M,\psi^*g)$  satisfy the  Einstein equations for the same matter distribution and are identical outside $H$. But they differ inside the hole. 
\end{itemize}
Einstein saw this as a proof that the matter distribution fails to determine the metric uniquely, and 
regarded this as such a severe challenge to determinism that, supported by the other problems he had at the time,  he  retracted general covariance.

 From a modern point of view the energy-momentum tensor is a red herring in the argument,\footnote{Continuing footnote \ref{EMach1}: Janssen (2007),  footnote 98, notes that Einstein formulated his
 requirement that the matter distribution fully determines the metric only in 1917; 
 in  1913  Einstein still thought of Mach's principle in the light of  the relativity of inertia. Furthermore, Einstein (1914)  explicitly introduced the final version of the hole argument in terms of a conflict
between general covariance and the ``law of causality'' (``\emph{Kausalgesetz}''), which was contemporary parlance for determinism. In sum, it seems safe to say, with Janssen (2007),  that the `worries
about determinism and causality that are behind Einstein's hole argument have strong Machian overtones.'
See Norton (1993) for Einstein's general struggle with general covariance, and its aftermath. }  which may just as well be carried out \emph{in vacuo}, as will be done from now on; this also strengthens my subsequent reformulation of the argument, since the theorem on which this is based is less well developed in the presence of matter. 

 Earman and  Norton (1987), then,  revived the \HA, as follows:\footnote{This (as well as Weatherall's critique) relies on a precise understanding of the notion of an  \emph{isometry} between spacetimes $(M',g')$ and $(M,g)$: this is a 
 diffeomorphism  $\psi:M'\raw M$ for which $g'=\psi^*g$, or, equivalently, $g=\psi_*g'$, where $\psi_*=(\ps\inv)^*$. In particular, following e.g.\ Hawking and  Ellis (1973), we always take an isometry to be a diffeomorphism.} 
\begin{enumerate}
\item Although, in the case considered by Einstein, $(M,\psi^*g)$ and $(M,g)$  are mathematically speaking different spacetimes (unless $\psi^*g=g$, in which case the \HA\ is void), 
physicists---usually tacitly---circumvent this alleged lack of determinism of \GR\ by simply ``identifying'' the two, i.e.\ by claiming that $(M,\psi^*g)$ and
$(M,g)$   represent ``the same physical situation''.
\item In this practice they are encouraged by the  observation that  $(M,\psi^*g)$  and $(M,g)$ are \emph{isometric}; trivially, the pertinent isometry is  $\ps$, and so the conclusion would be that
isometric spacetimes represent the same physical situation.
\item  \emph{However}---and this is their key point---this spells doom for spacetime substantivalists (like Newton), who (allegedly) should be worried 
that if in order to save determinism,  $x\in M$, carrying the metric $\psi^*g(x)$, must be identified with $\psi(x)\in M$, carrying the same metric,  then points have lost their ``this-ness'': they cannot be identified \emph{as such},  but only as carriers of  metric information. 
\item Thus one seems forced to choose between determinism and substantivalism.
\end{enumerate}
For the purpose of this paper, it is sufficient (and considerably easier) to replace the concept of substantivalism with what Gomes and  Butterfield (2023a) call \emph{Distinct}:\footnote{Gomes and  Butterfield (2023a)  argue that Earman and  Norton (1987) assumed the implication \emph{Substantivalism $\Raw$ Distinct}, which later literature questioned via attempts at ``sophistication''.}
\begin{center}
\emph{Though isometric, $(M,\psi^*g)$ and $(M,g)$ represent different physical possibilities.}
\end{center}
The tension exposed by the \emph{modern} \HA, then, is the one between \emph{Determinism} (cf.\ \S\ref{new}) and \emph{Distinct} (as opposed to general covariance, which is  assumed).\footnote{
Note that  there is a  kind of indeterminism in \GR\ that is outside the scope of the \HA\ (whatever its worth): In the language detailed in \S\ref{CBG} below, this is the possibility that strong cosmic censorship  (in the current, initial-value problem sense) fails;  in other words, that the \mghd\ (i.e.\ maximal Cauchy development) of some well-posed ``generic''  initial data for the Einstein equations is extendible in a suitable regularity class (of the metric).
See e.g.\  Dafermos (2019), 
 Doboszewski (2017, 2020),
     Smeenk and W\"{u}thrich (2021),  Landsman (2021), Chapter 10, and references therein.
     For conceptual history see also  Earman (1995) and  Landsman (2022). I return to this in  \S\ref{new}. \label{CSFN} }

 But this  discussion would be pointless if the \HA\ is a non-starter, as claimed by Weatherall (2018) and his followers (Fletcher, 2020; Bradley and   Weatherall, 2022; Halvorson and  Manchak, 2022).
Let me recall the main point:
\begin{small}
\begin{quote}
This discussion may be summed up as follows: There is a sense in which
$(M, g_{ab})$  and $(M, \til{g}_{ab})$ are the same, and there is a sense in which they are
different. The sense in which they are the same--that they are isometric, or
isomorphic, or agree on all invariant structure--is wholly and only captured
by $\til{\ps}$.
 The (salient) sense in which they are different--that they assign different
values of the metric to the same point--is given by an entirely different map,
namely, $1_M$. But--and this is the central point--one cannot have it both ways.
Insofar as one wants to claim that these Lorentzian manifolds are physically
equivalent, or agree on all observable/physical structure, one has to use $\til{\psi}$  to
establish a standard of comparison between points. And relative to this standard,
the two Lorentzian manifolds agree on the metric at every point--there is
no ambiguity, and no indeterminism. (This is just what it means to say that
they are isometric.) Meanwhile, insofar as one wants to claim that these
Lorentzian manifolds assign different values of the metric to each point,
one must use a different standard of comparison. And relative to this standard--that given by $1_M$--the two Lorentzian manifolds are not equivalent.
One way or the other, the hole argument seems to be blocked.\\ \mbox{} \hfill (Weatherall, 2018, p.\ 338--339)
 \end{quote}
\end{small}
 Here $\til{g}=(\psi\inv)^*g$. Furthermore, the notation $\til{\psi}$ stands for  the promotion $$(M, g)\stackrel{\til{\ps}}{\raw} (M, \til{g})$$  of the diffeomorphism $\ps:M\raw M$  of (bare) manifolds to an isometry of \emph{Lorentzian} manifolds, seen as the objects in a category  $\mathbf{Lor}$ whose arrows are isometries.\footnote{In view of Theorem \ref{YCBG} below, within such reasoning one should optimally work in the category $\mathbf{ST}$ of spacetimes (see footnote \ref{4dst}), whose isomorphism are isometries \emph{preserving time orientation}.} There is no such extension $\til{1}_M: (M, g)\raw (M, \til{g})$ of the 
 identity $\til{1}_M$ of $1_M$,  since $\til{1}_M$ is not an isometry (unless $\ps$ happens to be one);\footnote{The emphasis Halvorson and  Manchak (2022) put in this context on their otherwise highly valuable Theorem 1 (see footnote \ref{ufn}) seems like flogging a dead horse. This theorem implies that a hole diffeomorphism of the kind envisaged by Einstein (1914) and  Earmanand  Norton (1987), and explicitly constructed by Muller (1995), cannot be an isometry (which, or so it is suggested, would be the only remaining hope for the \HA\ to work, accepting Weatherall's critique). But if it were, then $\psi^*g=g$  all across $M$ and the dilemma of having both $(M,g)$ and $(M,\psi^*g)$ as models with the same matter distribution or other initial data simply would not arise: both (naive) determinism and substantivalism would be safe in \GR:
   the \HA\ would be a dud.}  in   $\mathbf{Lor}$ only $\til{1}_M: (M,g)\raw (M,g)$ is defined. And this is exactly Weatherall's point: one cannot meaningfully identify $x\in M$ \emph{seen as a point in the spacetime $(M,g)$} with $x\in M$ \emph{seen as a point in a different spacetime $(M,\til{g})$}, in order to be able to say that $\til{g}_{ab}(x)\neq g_{ab}(x)$, which would launch the \HA. A similar point was made by Penrose:
   \begin{small}
\begin{quote}
The basic principles of general relativity---as encompassed in the term `the principle
of general covariance' (and also `principle of equivalence')---tell us that there is no
natural way to identify the points of one spacetime with corresponding spacetime
points of another.\footnote{Taken from the penultimate version of Gomes (2021a);  omitted, alas, from the final version.}
 (Penrose, 1996, p.\ 591)
 \end{quote}
\end{small}

   A simpler way to make the same \emph{mathematical} point, in the spirit of Weatherall's own abelian group example but closer to the mathematical structure of \GR,
would be to take pairs $(M,f)$ where $M$ is a manifold (or  just a set without further structure) and $f:M\raw\R$ is a smooth function (or just a function), perhaps interpreted as some physical scalar field.  The allowed maps between pairs $(M',f')$ and $(M,f)$, i.e.\ the analogues of isometries,  are those diffeomorphisms (or just bijections) $\ps: M'\raw M$ for which $f'=f\circ\psi$. Taking $M'=M$, Weatherall would undoubtedly say:
\begin{itemize}
\item  one can  send a pair $(x,f(x))\in (M,f)$ to $(\psi(x), f(x))\in (M,\ps^*f)$, since $$f(x)=(\psi^*f)(\psi(x));$$
\item  but one cannot send $(x,f(x))$ to $(x,f(\psi\inv(x))$, although the latter is a point in $(M,\ps^*f)$, since neither $1_M$ nor $\ps$ can accomplish this.
\end{itemize}
Since, on this view, one cannot compare $(x,f(x))$ with $(x,f(\psi\inv(x))$, one cannot relate $f(\psi\inv(x))$ to $f(x)$ \emph{at} $x$ (which is  deemed crucial for the \HA). 

There are also philosophical arguments against such ``trans-world identifications'', see e.g.\ Lewis (1986) and,  in connection with the \HA, Butterfield (1988, 1989) and Gomes and  Butterfield (2023b). However, Weatherall  explicitly tries to undermine the \HA\ by appealing to \emph{mathematical practice}:
\begin{small}
\begin{quote}
In contemporary mathematics, the relevant standard
of sameness for mathematical objects of a given kind is given by the mathematical
theory of those objects. In most cases, the standard of sameness for
mathematical objects is some form of isomorphism. (\ldots) mathematical models of a physical theory are only defined up to isomorphism, where the standard of isomorphism is given by the mathematical
theory of whatever mathematical objects the theory takes as its models.
One consequence of this view is that isomorphic mathematical models in
physics should be taken to have the same representational capacities. By
this I mean that if a particular mathematical model may be used to represent
a given physical situation, then any isomorphic model may be used to represent
that situation equally well. Note that this does not commit me to the view
that equivalence classes of isomorphic models are somehow in one-to-one
correspondence with distinct physical situations. But it does imply that if
two isomorphic models may be used to represent two distinct physical situations,
then each of those models individually may be used to represent both
situations.  (Weatherall, 2018, p.\ 331--332)
 \end{quote}
\end{small}
The structuralist approach suggested here implies that the \emph{specific nature} of  individual objects (in category theory) or models
(in model theory) cannot be used.\footnote{This suggestion was subsequently somewhat weakened in 
Wheaterall (2021) and is also 
challenged by e.g.\  Roberts (2020), Gryb and Th\'{e}bault (2022), and Pooley and Read (2022).}

Weatherall's arguments are controversial: see e.g.\ 
Arledge and  Rynasiewicz (2019), Roberts (2020), Pooley and  Read (2021),  Gomes (2021ab), and Gomes and  Butterfield (2023a) for criticism and discussion.\footnote{See also Menon and  Read (2023) for a valid critique of Halvorson and  Manchak (2022).} My own two pennies worth would be to say that Weatherall uses the notion of ``contemporary mathematics'' quite selectively: in many cases the specific nature of mathematical objects--as opposed to just their isomorphism class--\emph{is} used. Indeed, the very \emph{definition} of an isometry rests on the ability to put $\ps^*g$ at $x$, where originally there was $g(x)$, and no mathematician would have any qualms saying this is the same $x$.
Subsequently, to call $\ps$ an isometry one needs to 
ask whether \emph{or not} $\psi^*g(x)$ equals $g(x)$ \emph{at} $x$. More generally, \emph{defining} 
the usual action of a diffeomorphism on a tensor (field) $T$ (like the metric) puts $(\psi^*T)(x)$ at $x$ where previously there was $T(x)$ \emph{at} $x$ (perhaps it is worth noting that in giving such definitions, mathematicians essentially stick to a Newtonian absolute spacetime, in which the points $x$ are identifiable).
 With it,
 the Lie derivative becomes questionable; see also Gomes (2021a), \S 2.4, and Gomes and  Butterfield (2023ab).
 Furthermore, though indeed out of step with  contemporary mathematics, 
 all of the local coordinate-based definitions of tensors used in the past by Einstein (and even, both before and after the introduction of \GR, by \emph{mathematicians} like Ricci and Levi-Civita), while awkward as \emph{definitions}, remain  valid \emph{theorems} in modern differential geometry. Do these definitions and  theorems now become suspect? Finally, even if Weatherall (2018) were right, should mathematical practice really dictate the way physicists must interpret the mathematical objects they use? I would say the opposite: \emph{physical} practice should dictate the way mathematical objects are used, at least in the context of mathematical physics.\footnote{For example, Alexandre Grothendieck shaped current mathematical practice. But he famously refused to have anything to do with physics (because in his view physics led to nuclear weapons). }
 \section{The  Choquet-Bruhat--Geroch theorem}\label{CBG}
 In any case, the apparently controversial \HA\ is clarified by placing it in the context of an uncontroversial theorem due to Choquet-Bruhat--Geroch (1969)  on the existence and uniqueness of maximal globally hyperbolic solutions to the Einstein field equations. This is contained in Theorem \ref{YCBG} below.\footnote{The original source is Choquet-Bruhat and  Geroch (1969), who merely sketched a proof (based on Zorn's lemma, which they even had to use twice). Even the 800-page textbook by  Choquet-Bruhat (2009)
does not contain a proof of the theorem (which is Theorem XII.12.2); the treatment in Hawking and  Ellis (1973), \S 7.6,  is slightly more detailed but far from complete, too. 
 Ringstr\"{o}m (2009) is a book-length exposition of the theorem, but ironically his proof of Theorem 16.6, i.e., Theorem \ref{YCBG} above,
 is wrong; it  is corrected in Ringstr\"{o}m (2013), \S 23. A constructive proof was given by Sbierski (2016), which is streamlined and summarized in Landsman (2021), \S 7.6. }
Together with Penrose's work on the causal structure of spacetime, it is one of the pillars of mathematical relativity. In particular, all \pde-related work in \GR\ is based on it, including  current approaches to cosmic censorship (see footnote \ref{CSFN}). 
In the theorem, Einstein's somewhat obscure way of stating the initial value problem for his field equations (in which,
  inspired by Mach's principle, he implicitly took initial data outside a hole, essentially at infinity) is replaced by a version first discussed by Hilbert (1917), who took initial data on a spacelike slice.\footnote{Hilbert (1917) gave the first analysis of \GR\ from a \pde\ point of view. He  addresses the  indeterminism  of Einstein's equations, and also refers to Einstein (1914), but does not explicitly relate his analysis to the \emph{Lochbetrachtung}. See also Howard and Norton (1993) and  Brading and Ryckman (2018). For some history of the \pde\ approach to \GR\  see  Stachel (1992),  Choquet-Bruhat (2014),  and Ringstr\"{o}m (2015), summarized in Landsman (2021), \S 1.9. It is also possible to give initial data for the Einstein equations on a \emph{null} hypersurface (Penrose, 1963); see e.g.\ 
 Klainerman and  Nicol\`{o} (2003) for a detailed treatment. That would also lead to a version of the \HA.
 }  A Hilbert-style \HA\ may then be based on Proposition \ref{HHA} below), which comes out
  a special case of  Theorem \ref{YCBG}. The change from Einstein's choice of initial data to Hilbert's  has little influence on any (relevant) philosophical discussion, which I will therefore base on Theorem \ref{YCBG}.

 Since the \HA\ is closely related to the following central issues in the philosophy of \GR, it is unsurprising that the Choquet-Bruhat--Geroch  theorem clarifies those as well (in a way that can be separated from the \HA):
 \begin{enumerate}
\item Finding an appropriate notion of determinism for \GR;
\item Interpreting isometries in \GR\ as gauge symmetries. 
\end{enumerate}
Together with a reconsideration of Weatherall's critique of the \HA, the first issue will be taken up in \S\ref{new}, especially in the light of previous proposals by Butterfield (1987, 1988, 1989).
The second will be discussed in \S\ref{GCSRT}.

First, I review the theorem in question. It is the culmination of the
 initial-value approach to \GR, which is based on \pde-theory, and the following ideology:\footnote{Physicists would see this as the \textsc{adm} approach to \GR, as in Misner,Thorne, and  Wheeler (1973). But the mathematical literature developed almost independently, led by Choquet-Bruhat.} 
\begin{itemize}
\item \emph{All valid \textbf{assumptions} in \GR\ are assumptions  about  initial data}  $(\til{\Sg},\til{g}, \tilde{k})$. 
\end{itemize}
Such an initial data triple,  assumed smooth, is obtained by 
 equipping some $3d$ Riemannian manifold $(\til{\Sg},\g)$  with a second symmetric tensor $\til{k}\in \X^{(2,0)}(\til{\Sg})$, i.e.\ of the same ``kind'' as the 3-metric $\g$, 
such that  $(\til{\Sg},\g,\tilde{k})$ satisfies the vacuum constraints \begin{align}
 \til{R} 
-\Tr(\tilde{k}^2)+\Tr(\tilde{k})^2 =0; &&
\til{\n}_j\tilde{k}_i^j-\til{\n}_i\Tr(\tilde{k})=0. \label{MC0}
\end{align}
Here $\til{R}$ is the Ricci scalar on $\til{\Sg}$ for the Riemannian metric $\g$ and likewise $\tn$ is the unique Levi-Civita (i.e.\ metric) connection on $\til{\Sg}$ determined by $\g$ (so that $\tn\g=0$).
\begin{itemize}
\item \emph{All valid \textbf{questions} in \GR\ are  questions about ``the'' \mghd\ $(M,g,\iota)$ thereof. }
\end{itemize}
Among these questions, the one relevant to the \HA\  concerns the uniqueness of  $(M,g,\iota)$, whence the scare quotes around `the'. 
Roughly speaking, a \mghd\ (for  \emph{maximal globally hyperbolic development}) of $(\til{\Sg},\g,\tilde{k})$ is a maximal spacetime $(M,g)$ ``generated'' by these initial data via the Einstein equations, in that  $$\iota:\til{\Sg}\hookrightarrow M$$ injects $\til{\Sg}$ into  $M$ as a ``time slice'' on which the 4-metric $g$ induces the given 3-metric $\g$ and extrinsic curvature $\til{k}$.
In more detail,\footnote{See also the references in footnote \ref{ufn}, or
Landsman (2021), \S 7.6.  Tildes adorn $3d$ objects. }  a \emph{Cauchy development}  or \emph{globally hyperbolic development} of given initial data $(\til{\Sg},\g,\tilde{k})$ satisfying the constraints \er{MC0} is a triple $(M,g,\iota)$, where $(M,g)$ is a  spacetime
 that solves the vacuum Einstein equations $R_{\mu\nu}=0$ and $\iota$ is an injection  making $\iota(\til{\Sg})$ a spacelike Cauchy (hyper)surface in $M$ such that $g$ 
  induces these initial data on $\iota(\til{\Sg})\cong\til{\Sg}$, i.e.\ $\g=\iota^*g$ is the metric and 
   $\tilde{k}$ is the extrinsic curvature of $\til{\Sg}$,  induced by the embedding $\iota$ and the 4-metric $g$.\footnote{
   Let $N$ be the unique (necessarily timelike) future-directed normal vector field on $\iota(\til{\Sg})$ such that $g_x(N_x,N_x)=-1$.
   Then $\til{k}(X,Y)=-g(\nabla_XN,Y)$ defines the extrinsic curvature of $\iota(\til{\Sg})$.}
     It follows that $(M,g)$ is globally hyperbolic, since it has a Cauchy surface.
         
 This formulation of the (spatial) initial-value problem for the (vacuum) Einstein equations was an achievement by itself. In particular, it  circumvents the vicious circle one is forced into if one tries to find initial data for an already given spacetime (solving the Einstein equations); for it is part of the problem to find the latter from the given initial data, and hence one cannot give say $dg/dt(t=0)$ as initial data.
  
  However, the main achievement concerns the existence and uniqueness of  $(M,g,\iota)$, which  depends on a suitable notion of \emph{maximality} (as in the far simpler case of \textsc{ode}s, where in order to guarantee uniqueness the time interval on which the solution is defined should be maximal). 
This notion is also non-trivial, and tied to \GR. Namely: 
 \begin{itemize}
\item 
A  \emph{maximal Cauchy development} or  \emph{maximal globally hyperbolic development},
 acronym
\mghd, 
of given smooth initial data $(\til{\Sg},\g,\tilde{k})$,  satisfying the constraints \er{MC0}, is a  Cauchy development 
$(M,g,\iota)$  with the property that for any other  Cauchy development  = globally hyperbolic development $(M',g',\iota')$  of these same data there exists an embedding $\ps:M'\raw M$ that preserves time orientation, metric, and Cauchy surface as defined by $\iota$, i.e., one has 
\begin{align}
g'=\psi^*g; && \iota'=\psi\inv\circ\iota. \label{psg} 
\end{align}
 \end{itemize}
A \HA\ \`{a} la Hilbert (1917) then follows from a simple observation:\footnote{This construction also works if $U=J^-(\iota(\til{\Sg}))$, cf.\ Curiel (2018) and Pooley (2022). The `Gauge Theorem' of Earman and Norton (1987), p.\ 520, is similar in spirit but lacks the connection to the initial-value problem that is central here.
Both results of course follows from general covariance.}
\begin{proposition}\label{HHA}
Given some  \mghd\  $(M,g,\iota)$ of the initial data  $(\til{\Sg},\g,\tilde{k})$, seen as the spacetime under review, 
let $U$ be an open neighbourhood of $\iota(\til{\Sg})$ in $M$. Take a (time orientation preserving) diffeomorphism $\psi$ of $M$ that is the identity on $U$. 
Then  the triple 
$(M',g',\iota')$, where $M'=M$, $g'=\psi^*g$ (so that $g'=g$ within $U$), and $\iota'=\iota$,
with time orientation induced by  $\psi$,\footnote{\label{TOFN} Defining time orientation by (the equivalence class of) a global timelike vector field $T$ on $M$, so that some causal vector $X$ is future-directed iff $g(X,T)<0$, 
this means that $T'=\psi\inv_*T$.} is a \mghd\ of the same
initial data $(\til{\Sg},\g,\tilde{k})$. 
\end{proposition}
This supports  a decent version of the \HA. It is  superior to Einstein's and Earman and  Norton's formulation in that it has shaken off any implicit reference to Mach's principle and is closer to the usual initial value problem for hyperbolic \pde s, with initial data on a spacelike hypersurface.\footnote{With a special \GR\ twist, though:  the Einstein equations are not hyperbolic, but the six spatial ones are hyperbolic in a suitable gauge, in which the remaining four are elliptic constraints.} Note in this respect that  the open set $U$  is the analog of the \emph{complement} of Einstein's hole. The larger $U$ is, the stronger the potential challenge to determinism (since the ensuing spacetimes differ in the  \emph{complement} of $U$, which for Einstein is inside the hole and for us is away from the initial data Cauchy surface  $\iota(\til{\Sigma})$), but although  $U$ can be made arbitrarily thin (as long as it contains 
 $\iota(\til{\Sigma})$), it may as well be arbitrary large (\emph{idem dito}). Thus the logical strength of both versions of the \HA\ seems quite similar. 

 But! With respect to  Weatherall's (2018) critique, Proposition \ref{HHA}, seen as Hilbert's version of the  \HA, is not really different from Einstein's, since 
  it equally well starts from a \emph{diffeomorphism} $\psi$ of $M$ that only becomes an \emph{isometry} from  $(M,\ps^*g)$ to $(M,g)$ ``with hindsight''.  Although I disagree with this critique (see \S\ref{intro}), I intend to weaken it even further  via
a slight reformulation--and corollary of--the celebrated
 theorem of Choquet-Bruhat and  Geroch (1969):\footnote{Though rarely if ever mentioned, the isometry $\ps$ in the converse is unique. This can be shown by Proposition 3.62 in O'Neill (1983) or the equivalent
  argument in footnote 639 of Landsman (2021), to the effect that  an isometry  $\ps$ is  determined at least locally (i.e.\ in a convex nbhd of $x$) by its tangent map $\ps_x'$ at some fixed $x\in M'$. Take $x\in \iota'(\til{\Sg})$. Since $\psi$ in Theorem \ref{YCBG} is fixed all along $ \iota'(\til{\Sg})$ by the second condition in \er{psg} and since it also fixes the (future-directed) normal $N_x$ to $\iota'(\til{\Sg})$ by the first condition in \er{psg}, it is determined locally.  Theorem 1 in Halvorson and  Manchak (2022) then applies, which is a rigidity theorem for isometries going back at least to Geroch (1969), Appendix A (as Halvorson and  Manchak acknowledge). \label{ufn}}
  \begin{theorem}\label{YCBG}
 For each initial data triple  $(\til{\Sg},\g,\tilde{k})$  satisfying the constraints \er{MC0} there exists  a \mghd\
 $(M,g,\iota)$. Any  triple $(M',g',\iota')$ that arises from an isometry $(M',g')\stackrel{\psi}{\raw}(M,g)$
that preserves time orientation and satisfies  $\psi\circ\iota'=\iota$ (fixing the  Cauchy surface) is an \mghd\ of the same initial data.\footnote{At first sight only the second half of \er{psg} appears in the above theorem. But the first half is part of the definition of an isometry.} Conversely, all \mghd s of these data arise in this way, so $(M,g,\iota)$ is unique up to these specific isometries. 
 \end{theorem}
 \noindent  The easy first part incorporates Proposition \ref{HHA} as a special case. The difficult second part, which is the real thrust of the theorem, is a
  nontrivial \emph{converse} to the first.
  \section{Rethinking the \HA}\label{new}
  Like the Earman--Norton  \HA,\footnote{Recall that Einstein's \HA\ was meant to enforce a choice between determinism and general covariance. Theorem \ref{YCBG} is based on standard (generally covariant) \GR\ and hence this choice has already been made, leaving the dilemma highlighted by Earman and  Norton (1987).}
    Theorem \ref{YCBG} exposes the tension between:
    \begin{enumerate}
\item  \emph{Determinism}, in the precise version that the Einstein equations for given initial data have a  unique solution in the sense that triples $(M,g,\iota)$ and $(M',g',\iota')$ as in the statement of Theorem \ref{YCBG} are seen as different mathematical representatives of the \emph{same} physical situation (i.e., are ``physically identified''). 
\item \emph{Distinct}, in the sense that 
 triples $(M,g,\iota)$ and $(M',g',\iota')$ represent \emph{different} physical possibilities 
(although they are observationally indistinguishable).  
\end{enumerate}
All discussions of this tension (e.g.\ Butterfield, 1989; Curiel, 2018; Pooley, 2022; Gomes \& Butterfield, 2023a), which I will not review, remain relevant if we support the \HA\ by Theorem  \ref{YCBG} instead of Proposition \ref{HHA} or Einstein's construction. However, after this replacement
   options 1 and 2  do differ a little from before:
\begin{itemize}
\item Option 1 is, in the context of  Theorem \ref{YCBG},  a larger move  compared to the original \HA, since far more spacetimes are now  declared to be ``physically equivalent'': namely all triples  $(M',g',\iota')$ in its statement. But in return the thrust of choosing this option is strengthened:  regarding $(M,\ps^*g)$ and $(M,g)$ as physically equivalent for some specific Hole diffeomorphism $\psi$, as in Proposition  \ref{HHA}, merely restores determinism in a special case, whereas (the second part of) Theorem \ref{YCBG} gives us complete assurance (barring indeterminism caused by violations of strong cosmic censorship, see \S\ref{intro} and below).
\item Option 2, on the other hand, requires no more commitment than in the original \HA: if we do not even identify $(M,g)$ with $(M,\psi^*g)$ in  Proposition  \ref{HHA}, where the underlying manifolds are the same, then certainly we will not identify any of the more general triples $(M',g',\iota')$, where they are different.
\end{itemize}

Given its much better embedding in the mathematical (physics) literature, it seems considerably more difficult for Weatherall and his followers to redirect their critique of the (modern) \HA\ to Theorem \ref{YCBG}. Although I can't speak for them, here are, prophylactically, some options they might still invoke, with a reply:
\begin{description}
\item[(a)]  The part of Theorem \ref{YCBG} that is actually relevant to the \HA\ should be stated as follows,  assuming the existence of a `reference' \mghd\ $(M,g,\iota)$:

\emph{Any diffeomorphism $\ps:M'\raw M$ gives rise to another \mghd\ 
 $(M',g',\iota')$ of the same initial data $(\til{\Sg},\g,\tilde{k})$, where $g'=\psi^*g$ and $\iota'=\ps\inv\circ \iota$.\footnote{$M'$ acquires a time orientation from $M$ and $\psi$, which $\ps$ trivially preserves, cf.\ footnote \ref{TOFN}. }}

Weatherall's original arguments (from his 2018) then apply almost  \emph{verbatim} (and indeed, I would say they are even clearer in this more general context). 
 But I was very careful in stating Theorem \ref{YCBG} the way I did: all reference to ``pure'' diffeomorphisms has gone, and all spacetimes that occur in the theorem are related by isometries:
 it is either \emph{assumed} (in the first half) or \emph{concluded} (in the second half) that
$(M',g')\stackrel{\psi}{\raw}(M,g)$ is an isometry. No other maps are mentioned and no controversial comparisons need to be made.

 In so far as an appeal to `mathematical practice' is made, I would answer that few if any mathematicians would be sensitive to the difference between the above reformulation of the middle part of Theorem \ref{YCBG} and its earlier statement.
 \item[(b)] One might accept Theorem \ref{YCBG} as it stands, but somehow object to Proposition \ref{HHA} being a special case of it (for example because its construction mixes up diffeomorphisms and isometries). 
Although I would again doubt any such arguments, especially if they appeal to `mathematical practice', even if they were valid I would point out that Theorem \ref{YCBG} \emph{as a whole} raises the same dilemma as the \HA\ and indeed may be taken to \emph{be} the \HA\ (2.0). 
\item[(c)] One could reject Theorem \ref{YCBG}, for example because
its  \emph{proof}  (admittedly!) does use (even local) diffeomorphisms and pointwise comparisons of metrics. But, further to the discussion in the Introduction, if Weatherall \emph{et al.} would object to generally accepted proofs of theorems by top mathematicians, then their appeal to ``mathematical practice'' would once again be self-defeating.\footnote{It may be interesting to point out that some of the most beautiful theorems in category theory, such as Gelfand duality, have very ugly proofs involving all kinds of constructions that the final result sweeps under the carpet; see Landsman (2017), Appendix C, for this specific example.}
\end{description}

I now compare the notion of \emph{Determinism} (which is standard in mathematical relativity) used above with an  influential definition appearing in the philosophy of physics literature due to Butterfield (1987, 1989), which was specifically developed in the context of the \HA. In order to clarify the connection of his definition with Theorem  \ref{YCBG},
I first state a somewhat awkward weakening of this theorem: 
\begin{corollary}\label{Dm2}
If two globally hyperbolic spacetimes $(M,g)$ and $(M',g')$ contain Cauchy surfaces $\til{\Sg}\subset M$ and $\til{\Sg}'\subset M'$, respectively, which carry initial data $(\til{\Sg},\g,\tilde{k})$ and  $(\til{\Sg}',\g',\tilde{k}')$ induced by 
the 4-metrics $g$ and $g'$ on $M$ and $M'$, respectively,  where both  $(M,g)$ and $(M',g')$ are  maximal for these initial data, and there is a 3-diffeomor\-phism $\al:\til{\Sg}'\raw\til{\Sg}$ such that $\g'=\al^*\g$ and $\tilde{k}'=\al^*\tilde{k}$, then there exists an isometry $\beta:M'\raw M$ that preserves time orientation and restricts to $\al$ on $\til{\Sg}$.
\end{corollary}\noindent
This corollary is weaker than Theorem  \ref{YCBG}, for it lacks the existence claim of  $(M,g)$. 

In comparison, Butterfield's Definition \textbf{Dm2}  of determinism is as follows:
\begin{small}
\begin{quote}
A theory with models $( M,O_i)$ is $\mathbf{S}$-deterministic, where $\mathbf{S}$ is a kind of region that occurs in manifolds of the kind occurring in the models, iff: \\ 
\mbox{} \hspace{25pt} given any two models  $( M,O_i)$ and  $( M',O_i')$ containing regions $S$ and $S'$ of kind $\mathbf{S}$, respectively, and any diffeomorphism $\al$ from $S'$ onto $S$:\\
\mbox{} \hspace{25pt}
 if $\al^*(O_i)=O_i'$ on $\al(S')=S$, then: there is an isomorphism $\beta$ from $M'$ onto $M$ that sends $S'$ to $S$, \emph{i.e.} $\beta^*O_i=O_i'$ throughout $M'$ and $\beta(S')=S$.\\ \mbox{}\hfill (Butterfield, 1987, p.\ 29; 1989, p.\ 9)\end{quote}
\end{small}
Here it would clarify the situation to add the requirement that $\beta$ extends $\al$.\footnote{Butterfield (1987, 1989) emphasizes that $\beta$ need not extend $\al$ (his primed objects are our unprimed ones). However, his counterexamples are easily avoided by requiring that $\al$ is only defined on $S'$. It is clear from Butterfield and Gomes (2023a), end of \S 2.2.3, that Butterfield endorses my discussion.
Butterfield contrasts  \textbf{Dm2} with a Laplacian kind of definition of determinism  \textbf{Dm1} he attributes to Montague and Earman:
  `A theory with models $( M,O_i)$ is $\mathbf{S}$-deterministic, where $\mathbf{S}$ is a kind of region that occurs in manifolds of the kind occurring in the models, iff:  given any two models  $( M,O_i)$ and  $( M',O_i')$  and any diffeomorphism $\beta$ from $M'$ onto $M$, and any
  region $S$ of $M$ of kind $\mathbf{S}$: if $\beta(S)$ is of kind $\mathbf{S}$ and also $\beta^*O_i=O_i'$ on $S'$, then: $\beta^*O_i=O_i'$ throughout $M'$.' If we correct this similarly to  \textbf{Dm2}, Butterfield's point still stands: the \HA\ (in any version) shows that \GR\ violates  \textbf{Dm1}.
   See also  Belot (1995),  Melia (1999), and Pooley (2022) for a detailed analysis of similar definitions. Pooley's version of 
  \textbf{Dm2} is a bit more general and also applies to \GR: `Theory $T$ is deterministic just in case, for any worlds $W$ and $W'$ that are possible according to $T$, if the past of $W$ up to some timeslice in $W$ is qualitatively identical to the past of $W$ up to some timeslice in $W'$, then $W$ and $W'$ are qualitatively identical.' Apart from my complaint that also this definition assumes the existence of $W$ and $W'$ (instead of proving it), a definition like this  requires a sub-definition of what is meant by `qualitative', which Theorem \ref{YCBG} also takes care of.
  } 

To start with the good news: though not intended for that purpose, this definition is quite suitable for expressing the idea of determinism inherent in strong cosmic censorship (and its possible violation!), cf.\ 
footnote \ref{CSFN}. Indeed, define models (of \GR)  to be  spacetimes $( M,g)$  satisfying the (vacuum) Einstein equations, and take the regions  $\mathbf{S}$ to be  \mghd s of initial data $(\til{g},\til{k})$ posed on some partial Cauchy $\til{\Sg}$ surface in $M$. Then \GR\ is $\mathbf{S}$-deterministic  precisely if  
strong cosmic censorship holds. 
 Indeed: if  not, 
let $(M,g,\iota)$ be a \mghd\ of initial data  $(\til{\Sg},\g,\tilde{k})$ as in Theorem \ref{YCBG}, with $( M,g)$ the associated spacetime, and let $( M',g')$ be a proper extension thereof (which by definition exists if strong cosmic censorship fails).
Then \textbf{Dm2} fails.\footnote{ The Kerr and Reissner-Nordstr\"{o}m spacetimes are indeterministic in precisely this way, though strictly speaking, these spacetimes are usually not seen as indicating a violation of strong cosmic censorship 
in \GR\ as a whole since their initial data are not ``generic'' in some suitable sense. }
 
 However, the ``Laplacian'' context in which \textbf{Dm2}  was originally proposed suggests that the idea was to take 
 $S\subset M$ to be a time-slice in a spacetime $( M, g)$. In that case, for there to be any hope that \GR\ is deterministic even if strong cosmic censorship holds, 
 Butterfield's  definition should be amended in the following way: 
\begin{enumerate}
\item The class of models should be restricted to maximal globally hyperbolic solutions to the vacuum Einstein equations with intial data as in Theorem \ref{YCBG};
\item Either $\mathbf{S}$ should be a neighbourhood of a Cauchy surface in $M$ (as in the \HA), or the models $( M, g)$ should be triples $(M,g,\iota)$, in which case one should  add a  condition on the extrinsic curvature to  `$\al^*(g)=g'$ on $\al(S')=S$'.
\end{enumerate}
Granting strong cosmic censorship, \GR\ is then $\mathbf{S}$-deterministic by  Theorem  \ref{YCBG}.

Of course, definitions like Butterfield's \textbf{Dm2} or its close cousins are not sacrosanct. An anonymous referee very ingeniously suggested that \emph{Determinism} and \emph{Distinct} are compatible with each other \emph{and with Theorem \ref{YCBG}} 
in the following way:
\begin{center}
\emph{The initial data induced by
 the triple $(\til{\Sg},\til{g},\til{k})$ on $\iota(\tilde{\Sg})\subset M$ are taken to be \emph{distinct} from those induced by the same triple on $\iota'(\tilde{\Sg})\subset M'$ (as in the theorem). }
 \end{center}
 Consequently, this distinction should then also be made in the situation of Proposition \ref{HHA} (which after all is the special case of the general situation just addressed),
  where the isometric space-times $(M,g)$ and $(M',g')$ both extend the ``same'' Cauchy surface $\iota(\Sg)=\iota'(\Sg)$  carrying the ``same'' initial data  $(\til{g},\til{k})$. Indeed, if the initial data for $(M,g)$ and $(M',g')$ are different, no determinist would object to the ensuing dynamically evolved space-times being different; whereas, according to this new view, they were misled in believing that the initial data for the isometric but otherwise different space-times $(M,g)$ and $(M',g')$ were the same.
  
   This kind of thinking would clearly be at odds with \textbf{Dm2} and similar (\GR-adapted) ``Laplacian'' definitions of determinism. It would also affect the original \HA: for likewise, a way out of both Einstein's and Earman and Norton's \HA\ would  be to deny that the space-times outside the hole within $(M,g)$ and $(M,\psi^*g)$ are identical (whereas the \HA\ is based on their identification). This would move the discussion from possible (mis)identifications of points and metrics \emph{inside} the hole (or far \emph{outside} the Cauchy surface), where the metrics are \emph{different}, to  (mis)identifications \emph{outside} the hole (or \emph{inside} the Cauchy surface), where the metrics are the \emph{same}. Claiming that such identifications are wrong or  undefined would be even more radical than Weatherall's objection to the \HA\ (which concerns the region where the metrics are \emph{different}), but would equally well undermine it, in the sense that the tension or  contradiction between \emph{Determinism} and \emph{Distinct} does not in fact arise. 
   
 This idea is certainly worth further discussion; for the moment, my reply is:
 \begin{description}
\item[-] No mathematical relativist would ever consider making the said distinction between initial data, since all their practices are based on  identifying them;
\item[-] Few philosophers would do so either, given the mismatch with  \textbf{Dm2} and the like.
\end{description}

 Finally, Theorem \ref{CBG} (as well as Theorem \ref{FTH} below) is reminiscent of the spontaneous breakdown of gauge symmetry through the Higgs mechanism.\footnote{See Struyve 2011,  Landsman (2017), \S 10.10, or any book on the Standard Model.}
Here, in order to settle into a minimum of the Higgs potential, the Higgs field $\phv$  must ``choose'' a point $\phv_c$ on a circle as its ``frozen'' vacuum value. The \emph{global} $U(1)$ symmetry involved in this choice is a \emph{finite-dimensional} shadow of the original \emph{infinite-dimensional local} $U(1)$  symmetry of the theory (see also footnote \ref{KNfn} below). Different choices of $\phv_c$ yield phenomenologically indistinguishable worlds and hence the analogy is between moving the vacuum value $\phv_c$ around on a circle and moving  a spacetime $(M,g)$ around in its orbit under its isometry group.\footnote{This analogy is admittedly weak, since Theorem  \ref{YCBG} involves both the embedding maps $\iota$ and the possibility that isometries move a given spacetime $(M,g)$ to one $(M',g')$ with a different underlying (but diffeomorphic) manifold $M$, neither of which have a counterpart in the Higgs mechanism.} Also here we are talking about symmetries of the universe as a whole, which is what makes them unobservable; the situation changes completely if different domains in the universe have different values of $\phv_c$. Let us now turn to an important special (!) case of this situation.
  \section{Special relativity: Status of the Poincar\'{e} group}\label{GCSRT}
  If one chooses the first option of \emph{Determinism} from the binary menu opening the previous section, 
  the isometries in Theorem \ref{YCBG} have to be interpreted accordingly as gauge  symmetries. The way these symmetries reflect the original diffeomorphism invariance of the Einstein equations (which invariance launched the \HA\ in the first place!) is clear from point \textbf{(a)} in the same section. In the special case $M'=M$,   for some given \mghd\ $(M,g,\iota)$ and any diffeomorphism $\psi:M\raw M$, one obtains a new
  \mghd\ $(M',g',\iota') =(M,\psi^*g,\psi\inv\circ\iota)$, which (by the vote for determinism) is deemed physically equivalent to $(M,g,\iota)$. There is no reason, however, why at least in the context of the initial-value formulation the diffeomorphism invariance of \GR\ should not include  diffeomorphisms $\psi:M'\raw M$ for $M'\neq M$ (and both mathematical and physical practice confirms this);  the statement that the ``gauge group'' of \GR\ is ``$\mathrm{Diff}(M)$'' seems too narrow (and even ill defined: which $M$ is meant?), and Theorem \ref{YCBG} shows the full situation. The seeming lack of general covariance of the initial-value \emph{problem} for \GR, notably of its initial data $(\til{\Sg},\til{g},\til{k})$,  is then amply compensated for by the fact that (by Theorem  \ref{YCBG})  its \emph{solution}, starting from some reference \mghd\ $(M,g,\iota)$, has covariance properties exceeding all  expectations:  
  the mathematical structure of the symmetry ``thing'' in the  initial-value formulation of \GR\ seems to be that of a \emph{groupoid}, where for given initial data $(\til{\Sg},\til{g},\til{k})$ the base space consists of all manifolds diffeomorphic to the  manifold $M$ in some reference \mghd\ $(M,g,\iota)$ of these data and the arrows are diffeomorphisms. The symmetry groupoid of \GR\ therefore depends on the initial data and is not universal.
  
On the other hand, the special case where $\psi$ is an isometry of $(M,g)$ (as opposed to a diffeomorphism promoted to an isometry from $(M,\psi^*g)$ to $(M,g)$, as in the \HA) is also clarified by Theorem  \ref{YCBG} (except for the identity map $1_M$, such isometries exist only for exceptional spacetimes). Indeed, since these are merely a special case of the general isometries in Theorem  \ref{YCBG}, they have to be interpreted in exactly the same way, that is, as gauge transformations. This may be surprising, because
   the isometry group $\mathrm{Iso}(M,g)$ of a fixed spacetime $(M,g)$ is finite-dimensional and hence is not given by freely specifiable functions on $M$,  as one expects in gauge theories.\footnote{  If $\dim(M)=n$, then for any semi-Riemannian metric $g$ the isometry group of $(M,g)$ is at most $\half n(n+1)$-dimensional.
 See O'Neill (1983), Lemma 9.28; Kobayashi and  Nomizu (1963), Theorem VI.3.3, do the Riemannian case.
Thus the Poincar\'{e}-group in $n=4$ has maximal dimension $10$.\label{KNfn} } The explanation is simply that $\mathrm{Iso}(M,g)$ is a finite-dimensional subgroup of the infinite-dimensional gauge groupoid just defined (for given initial data).\footnote{Subgroups of groupoids, seen as (small) categories in which each arrow is invertible (i.e.\ an isomorphism), are  contained in the group of  arrows from some base object to itself.}

The simplest case where this occurs in special relativity.\footnote{The following analysis  was inspired by correspondence with Henrique Gomes and Hans Halvorson, who  proposed to look at special relativity in this context. See also Iftime and Stachel (2006).}   Minkowski spacetime $\Mi=(\R^4,\eta)$ is a maximal globally hyperbolic solution to the vacuum Einstein equations,\footnote{\label{fn32} Maximality of Minkowski spacetime  follows from its inextendibility; see  e.g.\ Corollary 13.37 in O'Neill (1983) for the smooth case and 
 Sbierski (2018) for inextendibility even  in $C^0$.} which arises as ``the'' \mghd\ $(\R^4,\eta,\iota_0)$
 of the initial data $$(\til{\Sg}=\R^3, \til{g}=\dl,\til{k}=0),$$ where $\dl$ is the Euclidean metric on $\R^3$ and
 \beq
 \iota(x^1,x^2,x^3)=(0, x^1,x^2,x^3).\eeq 
  By the general analysis above, the group of  time orientation preserving Poincar\'{e} transformations is then contained in the gauge groupoid of these initial data and hence Poincar\'{e} transformations are, perhaps surprisingly, gauge transformations. 
  
 Another, more interesting way of reaching the same conclusion is to regard special relativity not as a specific solution to the vacuum Einstein equations, but as a generally covariant field theory by itself, formulated like \GR\ but with field equation 
\beq
R_{\rh\sg\mu\nu}=0,\label{Riem0}
\eeq
 instead of $R_{\mu\nu}=0$. 
 The initial value problem is then almost the same  as in general relativity,\footnote{This upsets the idea 
 that special relativity uses only linear subspaces of spacetime as hypersurfaces of simultaneity whereas general relativity uses general curved surfaces, but already Schwinger (1948) employed arbitrary initial data surfaces in  relativistic quantum field theory.}  
except that the 
 initial data  $(\til{\Sg},\g,\tilde{k})$ now satisfy the (vacuum) constraints
\begin{align}
\til{R}_{ijkl}-\til{k}_{il}\til{k}_{jk}+\til{k}_{ik}\til{k}_{jl}=0; &&
\tn_i\til{k}_{jk}-\tn_j\til{k}_{ik}&=0. \label{CRL5c}
\end{align}
The constraints  \er{CRL5c} of generally covariant special relativity are stronger than their counterpart \er{MC0} in \GR, which actually follows from  \er{CRL5c}
 by contracting with $\til{g}^{ik}\til{g}^{jl}$ and $\til{g}^{ik}$, respectively. The reason is that in \GR\ one merely asks for an embedding of the initial data  in a \emph{Ricci-flat} Lorentzian manifold $(M,g)$, i.e.\ $R_{\mu\nu}=0$, whereas in special relativity one seeks an embedding in a \emph{flat} Lorentzian manifold, as follows from 
 \er{Riem0} and the so-called fundamental theorem of (semi) Riemannian geometry.\footnote{In  Lorentzian signature this theorem states that $(M,g)$ is locally flat (in that its metric is locally Minkowski) iff its Riemann tensor vanishes. 
See e.g.\  Landsman (2021), Theorem 4.1.} 
 
 To avoid  global topological issues I assume that  $\til{\Sg}$ is diffeomorphic to $\R^3$, in which case the role of a (reference) \mghd\ $(M,g,\iota)$ in Theorem \ref{YCBG} is simply played by  Minkowski spacetime $(\R^4,\eta)$, with $\iota$ to be found (see Theorem \ref{FTH} below).\footnote{
 By the  splitting theorem of Geroch (1970) as improved by Bernal and S\'{a}nchez  (2003),
 global hyperbolicity of  $(M,g)$  gives $M\cong \R\x\til{\Sg}=\R^4$, diffeomorphically. Hence 
 we may actually take $M=\R^4$, due to  \er{Riem0} necessarily with the Minkowski metric. Finally, 
 $\Mi$ is  maximal, cf.\ footnote \ref{fn32}.}  
 One could now state and prove a counterpart of Theorem \ref{YCBG} for generally covariant special relativity, but instead I rely on 
 a Minkowskian version of the \emph{fundamental theorem for hypersurfaces}, whose original  version studied embeddings of two-dimensional surfaces $\Sg$ in $\R^3$ with Euclidean metric (here lies the origin of the Gauss--Codazzi equations, which also play a key role in deriving the constraints \er{MC0} in \GR):\footnote{See Kobayashi and  Nomizu (1969), Theorem VII.7.2 or  Landsman (2021), Theorem 4.18. This theorem is concerned with embeddings of curved surfaces with prescribed second fundamental form into Euclidean space and goes back to the nineteenth century. 
 The proof of the Minkowskian case is  the same, up to some sign changes: in the Euclidean case the first constraint in \er{CRL5c} is $\til{R}_{ijkl}+\til{k}_{il}\til{k}_{jk}-\til{k}_{ik}\til{k}_{jl}$, the sign changes going back to the different signs in the Gauss--Codazzi equations in Euclidean and Lorentzian signature, see e.g.\ eqs.\ (4.147) - (4.148)
in \S 4.7 in Landsman (2021). These sign changes do affect the outcome. For example, 
 Hilbert (1901) proved that it is impossible to isometrically embed two-dimensional hyperbolic space $(H^2,g_H)$  in  Euclidean $\R^3$. But hyperbolic space \emph{can} be isometrically embedded in $\R^3$ with Minkowski metric, cf.\ e.g.\  Landsman (2021), \S 4.4. Hence given $(H^2,g_H)$, a symmetric tensor $\tilde{k}$ such that  $(g_H,\tilde{k})$ satisfy the 
Euclidean constraint does not exist, but such a $\til{k}$ \emph{can} be found satisfying the Minkowski constraints. 
} 
  \begin{theorem}\label{FTH}
  For each initial data triple  $(\R^3,\g,\tilde{k})$  satisfying the constraints \er{CRL5c} there exists
  an isometric embedding  $\iota:\R^3\raw\R^4$ carrying the Minkowski metric $\eta$, whose extrinsic curvature is the given tensor $\til{k}$. Any triple $(\R^4, \eta, \iota')$ that arises from an  isometry $\psi$ of $\Mi$ (i.e.\ a Poincar\'{e} transformation), preserves time-orientation, and satisfies $\psi\circ\iota'=\iota$ 
has the same properties (that is, $\iota':\R^3\raw\R^4$ is an  isometric embedding and the extrinsic curvature induced on $\iota'(\R^3)\subset\R^4$ by the metric $\eta$  is $\til{k}$). 

Conversely, all triples $(\R^4, \eta, \iota')$ with these properties arise in this way from some given triple $(\R^4, \eta, \iota)$, which is therefore unique 
  up to Poincar\'{e} transformations.
\end{theorem}

There is a clear conceptual analogy between Theorems  \ref{YCBG} and \ref{FTH}, except that unlike the former, the latter does not take the spacetimes $(M,\eta')$ that are isometric to Minkowski spacetime $\Mi$ into account (where $M$ could even be $\R^4$). However, the corresponding more general version of Theorem \ref{FTH} would not affect my conclusion about 
  Poincar\'{e} transformations;  it would just assign a similar interpretation to even more  transformations.  And, exactly as in my discussion of special relativity as a special (vacuum) solution of \GR, this interpretation is that Poincar\'{e} transformations in generally covariant special relativity  play the same role as the isometries in general relativity that appear in Theorem \ref{YCBG}.  On the option of determinism as a way out of the \HA,  \emph{Poincar\'{e} transformations  are therefore physically inert}!
  
Now, whereas most physicists would be happy to regard isometries in general relativity 
as gauge symmetries, few  would regard Poincar\'{e} transformations as such. Fortunately, 
 Gomes (2021b), partly reflecting on Belot (2018), makes the right point:
\begin{small}
\begin{quote}
But some familiar symmetries of the whole Universe, such as velocity boosts in classical or
relativistic mechanics (Galilean or Lorentz transformations), have a \emph{direct} empirical significance
when applied solely to subsystems. Thus Galileo's famous thought-experiment about the ship---that a process involving some set of relevant physical quantities in the cabin below decks
proceeds in exactly the same way whether or not the ship is moving uniformly relative to the
shore--shows that sub-system boosts have a direct, albeit relational, empirical significance. For
though the inertial state of motion of the ship is undetectable to experimenters confined to
the cabin, yet the entire system, composed of ship and sea registers the difference between two
such motions, namely in the different relative velocities of the ship to the water.\\   \mbox{}\hfill (Gomes, 2021b, p.\ 2)
\end{quote}
\end{small}
In other words, 
in thinking about  Poincar\'{e} transformations as bringing physical change, as for example in boosts of Galileo's ship or Einstein's train, we apply such transformations to \emph{subsystems} of the universe. But Theorem \ref{FTH} concerns the action of  Poincar\'{e} transformations on spacetime \emph{as a whole}. See also Wallace (2022). Similarly, since time translations are Poincar\'{e} transformations, even special relativity seems a ``timeless'' theory in the sense that time translation is a gauge transformation. But once again, this only applies to empty spacetime, where it seems correct.

Summarizing: in the substantivalism versus relationalism debate  (Earman, 1989; Pooley,  2013)  I see  general relativity and  special relativity as qualitatively similar.  Whatever differences there are seem technical rather than conceptual, just reflecting the underlying difference between the field equations $R_{\mu\nu}=0$ and $R_{\rh\sg\mu\nu}=0$.
   \section{The \HA\ in the philosophy of science}\label{vF} 
Despite their denial of the \HA, Weatherall (2018) and Halvorson and  Manchak (2022) make some of the most pertinent comments towards its resolution:
\begin{small}
\begin{quote}
Mathematical models of a physical theory are only defined up to isomorphism, where the standard of isomorphism is given by the mathematical
theory of whatever mathematical objects the theory takes as its models.
One consequence of this view is that isomorphic mathematical models in
physics should be taken to have the same representational capacities. By
this I mean that if a particular mathematical model may be used to represent
a given physical situation, then any isomorphic model may be used to represent
that situation equally well. Note that this does not commit me to the view
that equivalence classes of isomorphic models are somehow in one-to-one
correspondence with distinct physical situations. But it does imply that if
two isomorphic models may be used to represent two distinct physical situations,
then each of those models individually may be used to represent both
situations. \\ \mbox{} \hfill (Weatherall, 2018, pp.\ 331--332)
\end{quote}
\end{small}
\begin{small}
\begin{quote}
Why is it, then, that there has been, and will surely continue to be, a feeling that there is some remaining open question about whether general relativity is fully deterministic? Our conjecture is that the worry here arises from the fact that general relativity, just like any other theory of contemporary mathematical physics, allows its user a degree of representational freedom, and consequently displays a kind of \emph{trivial semantic indeterminism}: how things are represented at one time does not constrain how things must be represented at later times.
\hfill (Halvorson and  Manchak, 2022, p. 19)
\end{quote}
\end{small}
These comments could just as well have been made about Theorem \ref{YCBG}, which by itself already makes it worth
 delving into the idea of ``representational freedom''.\footnote{See also Belot (2018), Fletcher (2020), Gomes (2021), Luc (2022),  and Pooley (2022).} 
 
I suggest that
the \HA\  and/or Theorem  \ref{YCBG} prompt us to choose not only between \emph{Determinism} and \emph{Distinct}, but, having chosen the first option, to also refine the consequences of this option--seeing isometries as gauge symmetries--through a further choice in this garden of forking paths. This second choice is
between two positions in the philosophy of mathematics that are traditionally seen as opposites, namely a Hilbert-style structuralism and a Frege-style abstractionism:\footnote{See e.g.\ Hallett (2010), Ebert and   Rossberg (2016), Mancosu (2016), Blanchette (2018),  Hellman and  Shapiro (2019), and Reck and  Schiemer (2020). Historically, Frege's abstractionism served his higher goal of logicism, but the former stands on its own and can  be separated from the latter. 
It may be objected that the heart of the Frege--Hilbert opposition does not lie in  abstractionism versus structuralism but in differences about the nature of mathematical axioms, definitions, elucidations, and existence, and in particular about Frege's insistence that every mathematical concept (such as ``point'' or ``line'') be defined on its own through reference, against Hilbert's revolutionary idea of implicit and ``holistic'' definition of concepts through an entire axiom system in which they occur. But these issues are closely related. For example, Hilbert's contextual and relational way of defining concepts naturally implies that whatever makes them concrete is given only up to isomorphism.
Abstractionism of the kind considered here arguably goes back to Aristotle, since the kind of equivalence relation lying at the basis of Frege's abstraction principle is typically obtained by Aristotle's procedure of \emph{abstraction by deletion}  (Mendell, 2019).
 For example, a mathematician sees a bronze sphere as a sphere, deleting its bronzeness. Also in so far as 
Hilbert famously claimed that mathematical objects exist as soon as the axioms through which they are implicitly defined are consistent (leaving their precise manner of existence in the dark, like Plato), the Frege--Hilbert opposition has its roots in the Aristotle--Plato one (Bostock, 2009). }
 \begin{itemize}
\item \emph{Structuralism:} spacetimes (with fixed initial data) are mathematical structures which by their very nature can only be studied up to isomorphism. Since isometry is the pertinent notion of isomorphism, the identifcation of isometric spacetimes called for by the \HA\ or Theorem  \ref{YCBG} was to be expected. 
\item \emph{Abstractionism:} the relevant mathematical object is the equivalence class of all spacetimes (with fixed initial data) up to isometry. Quoting Wilson (2010):
\begin{small}
\begin{quote}
Appeals to equivalence classes will seem quite natural if one
regards the novel elements as formed by \emph{conceptual abstraction} in
a traditional philosophical mode: one first surveys a range of concrete
objects and then \emph{abstracts} their salient commonalities. (Wilson, 2010, p.\ 395)
\end{quote}
\end{small}
In the case at hand, the `salient commonalities' seem to be the property that all members of a given equivalence class satisfy the vacuum Einstein equations with identical initial data. In the spirit of the abstractionist programme, this commonality may be expressed by the function $f$ from the class of all triples $(M,g,\iota)$ to the class of all triples $(\til{\Sg},\til{g},\til{k})$ that maps $(M,g,\iota)$ to the initial data it induces on $\iota(\Sg)\subset M$, where it is assumed that each $(M,g,\iota)$  is a 
maximal globally hyperbolic spacetimes with given Cauchy surface $\iota(\til{\Sg})$. 
\end{itemize}
These two options are put in perspective by the following quote from Martin, which Benaceraff chose as the opening quote of  his famous (1965):
\begin{small}
\begin{quote}
 \textsc{The} attention of the mathematician focuses primarily upon mathematical structure, and his intellectual delight arises (in part) from seeing that a given theory exhibits such and such a structure, from seeing how one structure is ``modelled'' in another, or in exhibiting some new structure and showing how it relates to previously studied ones
 \ldots But \ldots the mathematician is satisfied so long as he has some ``entities'' or ``objects'' (or ``sets'' or ``numbers'' or ``functions'' or ``spaces'' or ``points'') to work with, and he does not inquire into their inner character or ontological status. 
 
 The philosophical logician, on the other hand, is more sensitive to matters of ontology and will be especially interested in the kind or kinds of entities there are actually \ldots He will not be satisfied with being told merely that such and such entities exhibit such and such a mathematical structure. He will wish to inquire more deeply into what these entities are, how they relate to other entities \ldots Also he will wish to ask whether the entity dealt with is sui generis or whether it is in some sense \emph{reducible} to (or \emph{constructible} in terms of) other, perhaps more fundamental entities. 
 
 \mbox{} \hfill ---\textsc{R.M.~Martin}, \emph{Intension and Decision}
\end{quote}
\end{small}
Against abstractionism (both in the context of the \HA\ and in Frege's original application to the definition of Number), one may claim extravagance by noting that an equivalence class $[x]$ with respect to any equivalence relation $\sim$ on some given set $X$  is typically huge;\footnote{Recall that an equivalence class $[x]\subset X$ consists of all $y\in X$ such that $y\sim x$.}  no  theoretical or mathematical physicist ever works with such  equivalence classes of spacetimes, or even a tiny fraction of it.\footnote{See Gomes (2021a) for an analysis of physical practice, which in the context of gauge theories and \GR\ amounts to the  choice of  cross-sections of the canonical projection from $X$ to $X/\!\sim$.}

In practice, one picks some representative  $(M,g,\iota)$, from which one may switch to an equivalent triple  $(M',g',\iota')$ now and then, but one never uses the entire equivalence class.
And yet it is, strictly speaking, the entire equivalence class that Frege would invoke in order to obtain a proper definition or reference of the  word ``spacetime'' (provided the analogy with his definition of natural numbers is valid). 
See also Benaceraff (1965). 
To resolve this, one might try to work with the single object  $(\til{\Sg},\til{g},\til{k})$, i.e.\ the initial data that give rise to all of these isometric spacetimes, but no one does this either; all actual work in \GR\ is done in terms of just a few of the triples $(M,g,\iota)$, whose choice (within its isometry class) is made for convenience.

Within mathematical structuralism, the \HA\ seems compatible with both  structural realism (Ladyman, 2020) and  empiricist structuralism (van Fraassen, 2008); in the former, the structures in question are so to speak parts of reality whereas in the latter they model empirical phenomena. Let me quote van Fraassen:
\begin{small}
\begin{quote}
\begin{enumerate}
\item Science represents the empirical phenomena as embeddable in certain \emph{abstract structures} (theoretical models).
\item Those abstract structures are describable only up to structural isomorphism. 
\end{enumerate} (\ldots) 
How can we answer the question of how a theory or model relates to the phenomena by pointing to a relation between theoretical and data models, both of them abstract entities? The answer has to be that the data model represent the phenomena; but why does that not just push the problem [namely: \emph{what is the relation between the data and the phenomena it models}] one step back? The short answer is this:
construction of a data model is precisely the selective relevant depiction of the phenomena \emph{by the user of the theory} required for the possibility of representation of the phenomenon.  \\ \mbox{} \hfill(van Fraassen, 2008, pp.\ 238, 253)
\end{quote}
\end{small}
This last comment seems to describe the  practice of physicists and mathematicians working in \GR: some \emph{user of the theory} chooses a member $(M,g,\iota)$ of  its equivalence class, whilst some other \emph{user} (or even the same one) may pick another member.\footnote{Van Fraassen's  emphasis on the user also explains why say Kerr spacetime, even with fixed parameters $m$ and $a$, can be used to describe different black holes, despite the mathematical identity of the two models. Indeed, one user models the phenomena produced by one black hole, whilst another user uses (!) the ``spacetime'' in question to model the phenomena produced by another.}

In conclusion, empiricist structuralism seems to have strong cards in confronting  the \HA\ (in both its original versions or rephrased as Theorem \ref{CBG}): it does not suffer from the calculational intractability and  ontological extravagance of Frege-style abstractionism; and it seems to be warranted by actual scientific practice.
\begin{small}
\end{small}

\begin{thebibliography}{99}\addcontentsline{toc}{section}{References}
\bibitem{} Arledge, C., Rynasiewicz, R. (2019). On some recent attempted non-metaphysical dissolutions of the hole dilemma. \url{http://philsci-archive.pitt.edu/16343/}.
\bibitem{} Belot, G. (1995). New work for counterpart theorists: Determinism. \emph{The British Journal for the Philosophy of Science}  46, 185--195.
\bibitem{} Belot, G. (2018). Fifty million Elvis fans can't be wrong. \emph{Nous} 52, 946--981.
\bibitem{} Benaceraff, P. (1965). What numbers could not be. \emph{ Philosophical Review} 74, 47--73.
\bibitem{} Bernal, A. N.,   S\'{a}nchez, M. (2003). On smooth Cauchy hypersurfaces and Geroch's splitting theorem.  
\emph{Communications in Mathematical Physics} 243, 461--470.
\bibitem{} Blanchette, P. (2018). The Frege--Hilbert controversy. \emph{ Stanford Encyclopedia of Philosophy (Fall 2018 Edition)},  ed.\ Zalta, E.N. \url{https://plato.stanford.edu/archives/fall2018/entries/frege-hilbert/}.
\bibitem{}  Bostock, D. (2009). \emph{Philosophy of Mathematics} (Wiley).
\bibitem{} Brading, K.,  Ryckman, T. (2018). 
 Hilbert on general covariance and causality. \emph{Beyond Einstein. Einstein Studies Vol.\ 14}, eds.\
  Rowe D.E.,  Sauer, T., Walter, S.A., pp.\ 67--77 (Springer). 
\bibitem{} Bradley, C.,  Weatherall, J.O. (2022) Mathematical responses to the Hole Argument: Then and now.
\emph{Philosophy of Science} Volume 89,  1223--1232.
\bibitem{}  Butterfield, J. (1987). Substantivalism and determinism. \emph{International Studies in the Philosophy of Science} 2, 10--32. 
\bibitem{}  Butterfield, J. (1988). Albert Einstein meets David Lewis. \emph{PSA 1988}, 65--81.
\bibitem{}  Butterfield, J. (1989). The hole truth. 
  \emph{ British Journal for the Philosophy of Science} 40, 1--28. 
\bibitem{} Choquet-Bruhat, Y. (2009).
\emph{General Relativity and the Einstein Equations} (Oxford University Press).
\bibitem{}  Choquet-Bruhat, Y. (2014). Beginnings of the Cauchy problem. \url{https://arxiv.org/abs/1410.3490}.
\bibitem{} Choquet-Bruhat, Y.,  Geroch, R. (1969). Global aspects of the Cauchy problem in general relativity.
 \emph{Communications in Mathematical Physics} 14, 329--335.
  \bibitem{} Curiel, E. (2018). On the existence of spacetime
structure.      \emph{ British Journal for the Philosophy of Science} 69, 447--483.
  \bibitem{} Dafermos, M. (2019).
 The cosmic censorship conjectures in general relativity  (ICTP School on Geometry and Gravity). 
  Lecture 1:  \verb#https://www.youtube.com/watch?v=Lg1Cetf7V9I#. 
Lecture 2:  \verb#https://www.youtube.com/watch?v=SoRhBSt_mN0#.
 \bibitem{}  Doboszewski, J. (2017).
  Non-uniquely extendible maximal globally hyperbolic spacetimes in classical general relativity: A philosophical survey.
\emph{European Studies in Philosophy of Science} 6, 193-- 212.
 \bibitem{} Doboszewski, J. (2020). Epistemic holes and determinism in classical general relativity.
  \emph{ British Journal for the Philosophy of Science} 71, 1093--1111. 
    \bibitem{}  Earman, J. (1986). \emph{A Primer on Determinism} (Springer). 
  \bibitem{}  Earman, J. (1989). \emph{World Enough and Space-Time: Absolute versus Relational Theories of Space and Time} (The MIT Press).
  \bibitem{}     Earman, J. (1995).  \emph{Bangs, Crunches, Whimpers, and Shrieks: Singularities and Acausalities in Relativistic Spacetimes} (Oxford University Press).
 \bibitem{} Earman, J., Norton, J.D. (1987). What price substantivalism? The hole story.
     \emph{ British Journal for the Philosophy of Science} 9, 251--278. 
     \bibitem{} Ebert, P., Rossberg, M., eds. (2016).
     \emph{Abstractionism: Essays in Philosophy of Mathematics} (Oxford University Press).
           \bibitem{}    Einstein, A. (1914). Die formale Grundlage der allgemeinen Relativit\"{a}tstheorie.
\emph{Sitzungsberichte der K\"{o}niglich Preu\ss ischen Akademie der Wissenschaften (Berlin)} 1030--1085.
\url{https://einsteinpapers.press.princeton.edu/vol6-doc/100}.
           \bibitem{} Einstein, A. (1916). Die Grundlage der allgemeinen Relativit\"{a}tstheorie. \emph{Annalen der Physik} (4. S.) 49, 769--822. \url{https://einsteinpapers.press.princeton.edu/vol6-doc/311}.
\bibitem{}  Fletcher, S.C. (2020).  On representational capacities, with an application to general relativity.
\emph{Foundations of Physics} 50, 228--249. 
     \bibitem{}  Geroch, R. (1969). Limits of spacetimes.
 \emph{Communications in Mathematical Physics} 13, 180--193. 
      \bibitem{}  Geroch, R. (1970). Domain of dependence. \emph{Journal of Mathematical Physics} 11, 437--449.
        \bibitem{} Giovanelli, M. (2021).  Nothing but coincidences: The point-coincidence and Einstein's struggle with the meaning of coordinates in physics. \emph{European Journal for Philosophy of Science} 11:45.
      \bibitem{}  Gomes, H. (2021a). \emph{Why gauge? Conceptual Aspects of Gauge Theories.}
      PhD Thesis, University of Cambridge. \url{https://arxiv.org/abs/2203.05339}. 
          \bibitem{}  Gomes, H. (2021b).    Holism as the empirical significance of symmetries.
          \emph{European Journal for Philosophy of Science} 11:87. 
            \bibitem{}  Gomes, H. (2022a). Same--diff? Conceptual similarities between gauge transformations and diffeomorphisms. Part I: Symmetries and isomorphisms. \url{https://arxiv.org/abs/2110.07203}.
             \bibitem{}  Gomes, H. (2022b). Same--diff? Conceptual similarities between gauge transformations and diffeomorphisms. Part II:  Challenges to sophistication. 
  \url{https://arxiv.org/abs/2110.07204}.                 
        \bibitem{} Gomes, H., Butterfield, J. (2023a).
          The Hole Argument and beyond, Part I: The story so far. \url{https://arxiv.org/pdf/2303.14052.pdf}. 
          \bibitem{} Gomes, H., Butterfield, J. (2023b).
          The Hole Argument and beyond, Part II:  Treating non-isomorphic spacetimes.
          \url{https://arxiv.org/pdf/2303.14060.pdf}. 
           \bibitem{}  Gryb, S., Th\'{e}bault, K.P.Y. (2022), Regarding the `Hole Argument' and the `Problem of Time'.  \emph{Philosophy of Science} 83, 563--584. 
         \bibitem{} Hallett, M. (2010). Frege and Hilbert. \emph{ Cambridge Companion to Frege}, eds.\ 
   Potter, M., Ricketts, T., 
   pp.\ 413--464 (Cambridge University Press).   
\bibitem{} Halvorson, H., Manchak, J.B. (2022). Closing the Hole Argument. \emph{ British Journal for the Philosophy of Science}, in press. \url{http://philsci-archive.pitt.edu/19790/}.
\bibitem{} Hawking, S.W., Ellis, G.F.R. (1973). \emph{The Large Scale Structure of Space-Time} (Cambridge University Press). 
\bibitem{}  Hellman, G., Shapiro, S. (2019). \emph{Mathematical Structuralism} (Cambridge University Press). 
\bibitem{} Hoefer, C. (1994). Einstein's struggle for a Machian gravitation theory. \emph{Studies in History and Philosophy of Science} 25, 287--335.
\bibitem{}  Hilbert, D. (1901). \"{U}ber Fl\"{a}chen von Constanter Gaussscher Kr\"{u}mmung. 
\emph{Transactions of the American Mathematical Society}
 2, 87--99.
\bibitem{}  Hilbert, D. (1917). Die  Grundlagen der Physik (Zweite Mitteilung). 
 \emph{Nachrichten von der K\"{o}niglichen Gesellschaft der Wissenschaften zu G\"{o}ttingen, Mathematisch-Physikalische Klasse}, 53--76. 
 \bibitem{}  Howard, D.,  Norton, J.D. (1993).
 Out of the labyrinth? Einstein, Hertz, and the G\"{o}ttingen answer to the hole argument.
 \emph{The Attraction of Gravitation: New Studies in the History of General Relativity, Volume 5},
 eds.\  Earman, J.,  Janssen, M.,  Norton, J.D., pp.\  30--62 (Birkh\"{a}user).
 \bibitem{} Iftime, M.,  Stachel, J. (2006). The hole argument for covariant theories. \emph{General Relativity
and Gravitation} 38, 1241--1252.
 \bibitem{} Janssen, M. (2007). What did Einstein know and when did He know it? A Besso memo dated August 1913. \emph{The Genesis of General Relativity, Volume 2}, ed. Renn, J.,  pp.\ 787--837 (Springer). 
\bibitem{} Janssen, M., Renn, J. (2022). \emph{How Einstein Found His Field Equations:
Sources and Interpretation} (Springer). 
\bibitem{} Klainerman, S.,  Nicol\`{o}, F.  (2003). \emph{The Evolution Problem in General Relativity} (Birkh\"{a}user).
\bibitem{}Kobayashi,  S.,   Nomizu, K. (1963).
\emph{Foundations of Differential Geometry. I.} (Wiley). 
\bibitem{}Kobayashi,  S.,   Nomizu, K. (1969).
\emph{Foundations of Differential Geometry. II. } (Wiley). 
\bibitem{} Ladyman, J. (2020). Structural Realism. \emph{ Stanford Encyclopedia of Philosophy (Winter 2020 Edition)}, ed.\ Zalta, E.N. \url{https://plato.stanford.edu/archives/win2020/entries/structural-realism/}.
\bibitem{} Landsman, K. (2017). \emph{Foundations of Quantum Theory: From Classical Concepts to Operator Algebras} (Springer).  Open Access from \url{https://link.springer.com/book/10.1007/978-3-319-51777-3}.
\bibitem{} Landsman, K. (2021). \emph{Foundations of General Relativity: From Einstein to Black Holes} (Radboud University Press). Second corrected and expanded printing.  \url{https://radbouduniversitypress.nl/site/books/m/10.54195/EFVF4478/}.
 \bibitem{} Landsman, K. (2022). Penrose's 1965 singularity theorem: From geodesic incompleteness to cosmic censorship.
\emph{General Relativity and Gravitation} 54:115 (2022).
\url{https://link.springer.com/article/10.1007/s10714-022-02973-w}. 
\bibitem{} Lewis, D. (1986). \emph{On the Plurality of Worlds} (Blackwell). 
\bibitem{} Luc, J. (2022). Arguments from scientific practice in the debate about the physical equivalence of symmetry-related models. \emph{Synthese}  200:72.  \url{https://doi.org/10.1007/s11229-022-03618-w}.
\bibitem{} Mancosu, P. (2016). \emph{Abstraction and Infinity} (Oxford University Press).
\bibitem{} Maudlin, T. (1990). Substances and spacetime: What Aristotle would have said to Einstein. \emph{Studies in History and Philosophy of Science} 21, 531--561.
\bibitem{} Melia, J. (1999). 
Holes, haecceitism and two conceptions of determinism.
\emph{The British Journal for the Philosophy of Science} 50, 639--664.
\bibitem{} Mendell, H. (2019). Aristotle and Mathematics. \emph{ Stanford Encyclopedia of Philosophy (Fall 2019 Edition)},  ed.\ Zalta, E.N. \url{https://plato.stanford.edu/archives/}  \url{fall2019/entries/aristotle-mathematics}.
\bibitem{} Menon, T., Read, J. (2023). Some remarks on recent mathematical-\emph{cum}-formalist responses to the \HA. Unpublished preprint. 
\bibitem{}  Misner, C.W., Thorne,  K.S., Wheeler, J.A. (1973). \emph{Gravitation} (Freeman).
\bibitem{} Muller, F.A. (1995). Fixing a hole. \emph{Foundations of Physics Letters} 8, 549--562.
\bibitem{} Norton, J.D. (1993). General covariance and the foundations of general relativity: Eight decades of dispute.
\emph{Reports on Progress in Physics} 56, 791--858. 
\bibitem{} Norton, J.D. (2019).  The Hole Argument. \emph{ Stanford Encyclopedia of Philosophy (Summer 2019 Edition)}, ed.\ Zalta, E.N. \url{https://plato.stanford.edu/archives/sum2019/entries/spacetime-holearg/}.
\bibitem{} 
 O'Neill, B. (1983). \emph{Semi-Riemannian Geometry} (Academic Press). \bibitem{} Penrose, R. (1963).  Null hypersurface initial data for classical fields of arbitrary spin and for general relativity.
 \emph{Aerospace Research Laboratories} 63--65. Reprinted in \emph{General Relativity and Gravitation} 12, 225--264 (1980).
\bibitem{}  Penrose, R. (1979). Singularities and time-asymmetry. \emph{General Relativity: An Einstein Centenary Survey},
        eds.\  Hawking, S.W.,  Israel, W., pp.\  581--638  (Cambridge University Press). 
\bibitem{} Penrose, R. (1996). On gravity's role in quantum state reduction. \emph{General Relativity
and  Gravitation} 28, 581--600.        
\bibitem{} Pooley, O. (2013). Substantivalist and relationalist approaches to spacetime.  \emph{ Oxford Handbook of Philosophy of Physics}, ed.\ R. Batterman, chapter 16,  (Oxford University Press). 
\url{http://philsci-archive.pitt.edu/9055/}.
\bibitem{} Pooley, O. (2022). The hole argument. \emph{ Routledge Companion to Philosophy of Physics}, eds.\ Knox, E., Wilson, A., chapter 10 (Taylor and  Francis).  \url{http://philsci-archive.pitt.edu/18142/}.
\bibitem{} Pooley, O.,  Read, J. (2021). On the mathematics and metaphysics of the Hole Argument.
 \emph{ British Journal for the Philosophy of Science}, in press. \url{https://doi.org/10.1086/718274}.
\url{http://philsci-archive.pitt.edu/19774/}.
\bibitem{} Reck, E.,   Schiemer, G. (2020). Structuralism in the philosophy of mathematics. \emph{ Stanford Encyclopedia of Philosophy (Spring 2020 Edition)}, ed.\ Zalta, E.N.  \url{https:plato.stanford.edu/archives/spr2020/entries/structuralism-mathematics/}. 
\bibitem{}  Ringstr\"{o}m, H. (2009). \emph{The Cauchy Problem in General Relativity} (EMS). \bibitem{} 
Ringstr\"{o}m, H. (2013). \emph{On the Topology and Future Stability of the Universe}
(EMS). 
\bibitem{}   Ringstr\"{o}m, H. (2015). Origins and development of the Cauchy problem in general relativity.
\emph{Classical and Quantum Gravity} 32, 124003.
   \bibitem{}  Roberts, B.W. (2020). Regarding `Leibniz equivalence'. \emph{Foundations of Physics}
   50, 250--269. 
   \bibitem{} Sbierski, J. (2016). On the existence of a maximal Cauchy development for the Einstein equations: A dezornification.
\emph{Annales Henri Poincar\'{e}} 17, 301--329. 
   \bibitem{} Sbierski, J. (2018). The $C^0$-inextendibility of the Schwarzschild spacetime and the spacelike diameter in Lorentzian geometry. \emph{Journal of Differential Geometry}  108, 319--378. 
\bibitem{} Schwinger, J. (1948). Quantum electrodynamics. I. A covariant formulation.
\emph{Physical Review} 74, 1439--1461.
\bibitem{} Smeenk, C.,  W\"{u}thrich, C. (2021).  Determinism and general relativity.
\emph{Philosophy of Science} 88, 638--664.
   \bibitem{}  Stachel, J. (1992). The Cauchy problem in general relativity--The early years.   
   \emph{Studies in the History of General Relativity},  eds.\ Eisenstaedt, J.,  Kox, A.J.,
   pp.\ 407--418 (Birkh\"{a}user). 
   \bibitem{}   Stachel, J. (2014).  The hole argument and some physical and philosophical implications.
   \emph{Living Reviews in Relativity} 17, 1-66.  \url{https://link.springer.com/article/10.12942/lrr-2014-1}.
           \bibitem{} Struyve, W. 2011. ``Gauge invariant accounts of the Higgs mechanism.'' { Studies in History and Philosophy of Modern Physics} 42, 226--236.
     \bibitem{}  Van Fraassen, B.C. (2008). \emph{Scientific Representation} (Oxford University Press).  
          \bibitem{} Wallace, D. (2022). Isolated systems and their symmetries, part II: Local and global symmetries of field theories. \emph{Studies in History and Philosophy of Science} 92, 249--259.
   \bibitem{}   Weatherall, J.O. (2018).  Regarding the `hole argument'.
      \emph{ British Journal for the Philosophy of Science}  69,  329--350. 
         \bibitem{}   Weatherall, J.O. (2021). 
          Why not categorical equivalence? \emph{Hajnal Andr\'{e}ka and Istv\'{e}n N\'{e}meti on Unity of Science: From Computing to Relativity Theory Through Algebraic Logic}, eds.\   Madar\'{a}sz, J.,  Sz\'{e}kely, G., 
           pp.\ 427--451 (Springer).
      \bibitem{} Wilson, M. (2010). Frege's mathematical setting.  
         \emph{ Cambridge Companion to Frege}, eds.\   Potter, M., Ricketts, T., 
   pp.\ 379--412 (Cambridge University Press).    
\end{thebibliography}
\end{document}